\documentclass{article}
\usepackage{epsfig}
\usepackage{graphicx}
\usepackage{leftidx}
\def \be{\begin{equation}}
\def \ee{\end{equation}}
\def \bea{\begin{eqnarray}}
\def \eea{\end{eqnarray}}
\begin{document}
\begin{center}
{\bf \large Signals of additional $Z'$ boson in $e^+e^-\to W^+W^-$
at the ILC with polarized beams  
}
\vskip 1cm
{\bf B. Ananthanarayan\\
Monalisa Patra}\\
\medskip
Centre for High Energy Physics\\
Indian Institute of Science\\
Bangalore 560 012, India\\
\bigskip
{\bf P. Poulose}\\
\medskip
Department of Physics\\
Indian Institute of Technology Guwahati\\
Assam 781 039, India\\
\bigskip

\vskip 2cm

\end{center}
\begin{abstract}
We consider the possibility of fingerprinting the presence of 
heavy additional $Z'$ bosons that arise naturally in extensions of the 
standard model such as $E_6$ models and left-right symmetric models, through 
their mixing with the standard model $Z$ boson.  
By considering a class of observables 
including total cross sections, energy distributions and angular 
distributions of decay leptons we find significant deviation from 
the standard model 
predictions for these quantities with right-handed electrons and 
left-handed positrons at $\sqrt{s}$=800 GeV.  The deviations being less 
pronounced at smaller centre of mass energies as the models are already 
tightly constrained.  Our work suggests that the ILC should have a 
strong beam polarization physics program particularly with these 
configurations.  On the other hand, a forward backward asymmetry
and lepton fraction in the backward direction 
are more sensitive to new physics with realistic polarization
due to interesting interplay with the neutrino t- channel diagram.
This process complements the study of fermion pair production
processes that have been considered for discrimination
between these models.

\end{abstract}

\newpage 
\section{Introduction}
The International Linear Collider (ILC) is a proposed high
energy, high luminosity electron-positron collider with
the mission of studying the standard model (SM) at high precision
and to look for signals beyond the standard model~\cite{ILC}. 
It has been proposed that an initial beam polarization program
can also significantly enhance its capabilities in meeting
these objectives, see ref.~\cite{review}.

One of the important processes that will be studied at high
precision at ILC with and without beam 
polarization~\cite{review} is $W$-pair production.  
Phenomenological studies of this process within the 
SM and
some extensions have been carried out in great detail 
~\cite{Hagiwara,Gounaris} starting many years ago. 
Since properties of the weak gauge bosons are closely linked to electroweak 
symmetry breaking (EWSB) and the structure of the gauge sector 
in general, detailed study of $W$ physics
will throw light on what lies beyond the SM.  

On the other hand, it is entirely likely that there 
are additional $Z$ bosons, denoted by $Z'$ in the TeV range, see for
instance the review section in ref.~\cite{RPP}.  These are
present in several economical extensions of the standard
model.  With this strong motivation,
signatures of such a gauge boson is searched for in the past and existing 
colliders. Direct and indirect searches at LEP as well as at TeVatron and other
existing facilities provide bounds on the masses of this particle and on other
model parameters.  Direct searches at TeVatron put lower limits of 
$630 - 1030$ GeV \cite{Aaltonen:2008ah,Jaffre:2009dg} and LEP 2 put limits
of $673 - 1787$ GeV \cite{Alcaraz:2006mx}, 
while electroweak precision analysis of LEP provide lower limits of 
$475 - 1500$ GeV \cite{Erler:2009jh} on the mass, depending on the model considered. 
Even if not directly produced, they can be
finger printed easily as they would mix with the traditional
$Z^0$ of the electroweak model.  Thus $W$-pair production process has 
a winning edge compared to fermion pair production when it comes to the effect
of this mixing. This is because, as these $Z'$ do not interact
with the standard $W$ bosons, the $W$-pair production process is insensitive to
the presence of $Z'$ in the absence of mixing. 
In contrast, fermion pair production
process is sensitive to the presence of $Z'$ even in the absence of mixing.

We note that although such mixing is highly constrained 
by precision measurements at LEP and by other existing experimental 
data~\cite{RPP},
with the high statistics expected at the ILC for W-pair production, 
even such small mixing can be probed effectively. 
The new effects could be manifested in departures from the standard model
cross section for W-pair production, and in various
differential distributions and asymmetries.  

Recently in the context of the littlest Higgs model(LHM), which also
contain $Z'$ bosons, we
demonstrated the utility of several simple distributions
in fingerprinting the model~\cite{APP}.  As a result, it may be worthwhile
considering how other popular $Z'$ models arising
in $E_6$ unification and so called left-right symmetric models (LRSM) and
alternative left-right symmetric models (ALRSM),  which have also
been considered by the CLIC Physics Working Group~\cite{CLIC},
can subject themselves
to a diagnosis.  This is the main aim of the present work.
We have studied the process at reference energies of
500, 800 and 1000 GeV, and find that effects are pronounced
only at the higher energy.
In addition, we conclude that initial beam polarization 
can significantly enhance the diagnostic ability of the ILC.
In particular, for the class of models we are considering
these effects are of importance for right-handed electrons and
left-handed positrons in the process we are considering.  This may be understood 
as arising from the dominant $t$-channel SM contribution which gets erased
away with this choice of beam polarization revealing the new physics
contributing through the $s$-channel. The above argument holds in the case
for realistic degrees of polarization of the beams;
it is expected that about 90\% electron polarization would be
achievable along with a positron polarization of 60\%~\cite{review}. We present
our results for both these cases and find that there are interesting
effects even for the latter case as the t- channel contribution which now
survives, can play an interesting and effective role.

We will use the observables considered by us in 
our earlier work, ref.~\cite{APP}, which are total
cross sections, single energy
distribution of the secondary lepton, 
lepton angular distribution and forward-backward (FB)
and left-right (LR) asymmetries.  In addition to the above we
consider an important and useful energy-energy correlation
of the type first considered in the context of some anomalous
gauge couplings by Dicus and Kallianpur~\cite{DK}.  Despite its obvious
utility it has not received much attention, and we will demonstrate how this correlation
in combination with beam polarizations can extract important information on $Z'$ models.

The scheme of this paper is the following:  In Sec.~\ref{Models} we 
discuss the details of the models we are considering.  In Sec.~\ref{Analysis}
we consider the kinematics of the process and the subsequent decays
in great detail.  The section is organized in several subsections
for convenience.
In Sec.~\ref{Conclusions} we present a 
discussion including a comparison of $W$-pair
production with fermion pair production
considered in the literature, and our conclusions.

\section{$Z'$ Models}\label{Models}
The presence of an additional neutral gauge boson ($Z'$) is anticipated 
in many extensions of the SM. Some grand
unified theories (GUT) like those based on $E_6$, which contain
the gauge group of
$SU(3)_C\times SU(2)_{L}\times U(1)_Y\times U(1)'$ clearly have
additional $U(1)'$ symmetries (for a recent review, see,
e.g. ref.~\cite{Langacker}). These could also
arise in superstring theories.  There are many other extensions of the SM with 
dynamical symmetry breaking, Little Higgs models (LHM), LRSM and ALRSM
also with extended gauge sectors. In models with large extra
dimensions, Kaluza-Klein excitations of the SM gauge bosons propagating in
the bulk manifest as extra gauge bosons in four dimensions. 
For our purposes we confine our attention to certain GUT models
based on $E_6$ and LRSM and ALRSM.
We will study the presence of an extra $U(1)'$ 
symmetry present in addition to the SM gauge symmetries, 
arising in the candidate models mentioned above.
In some of these models there could be more than one neutral gauge 
bosons along with possible presence of heavy charged gauge bosons. 
We assume that any such additional gauge bosons 
decouple from the particle spectrum under study, and therefore will
be ignored in this study. 

The new gauge boson $Z'$ could mix
with the SM gauge boson to give the physical eigenstates. 
As explained later in this section, the $W$-pair production in $e^+e^-$ collisions  
has the advantage of directly probing the mixing unlike, 
for example, the fermion pair production process. 
This is because, the $W$ does not interact directly with the $Z'$.
While such mixing is highly constrained by the current experimental constraints, 
it may still be possible to probe its effect in a high energy, high luminosity
machine like the ILC.
With very high statistics expected for $W$-pair production at ILC, 
this process has the potential to probe even very small
mixing permitted.

Let us now turn to some general features of the scenario.
Here we closely follow the discussion in ref.~\cite{Langacker, PP}.
With one additional $U(1)'$ symmetry, the mass term of the 
neutral gauge bosons may be written as,
\be
L_Z^{\rm mass}=\frac{1}{2}
\left(\begin{array}{cc} Z^{0\mu}&Z'^\mu \end{array}\right)
\left(\begin{array}{lr} 
M_{Z^0}^2&\Delta^2\\[2mm]
\Delta^2& M_{Z'}^2 \end{array}\right)
\left(\begin{array}{c} Z^0_\mu\\Z'_\mu \end{array}\right)
\ee
Diagonalizing the above mass matrix the mass term is presented in terms of
the physical boson fields as
\be
L_Z^{\rm mass}=\frac{1}{2} M_1^2~Z_1^\mu Z_{1\mu}+\frac{1}{2} M_2^2~Z_2^\mu Z_{2\mu},
\ee
where we identify the lighter $Z_1$ as the observed $Z$-boson with 
$M_1=91.19$ GeV, and the $Z_2$ as its heavier counterpart. 
In terms of the SM gauge boson, $Z_0$ and the new gauge bosons,
$Z'$, we may write the physical states as
\be
\left(\begin{array}{c} Z_1\\Z_2 \end{array}\right)=
\left(\begin{array}{rr} 
\cos\theta&\sin\theta\\ 
-\sin\theta&\cos\theta \end{array}\right)
\left(\begin{array}{c} Z^0\\Z' \end{array}\right).
\ee
The mixing angle, $\theta$ is related to the diagonalization of the mass
matrix, and can be expressed in terms of the physical masses and the SM
mass parameter as
\be
\tan^2\theta=\frac{M_{Z^0}^2-M_1^2}{M_2^2-M_{Z^0}^2}.
\label{eqn_tanb}
\ee 
For our phenomenological analysis it is more convenient to 
re-parametrize the above 
by defining a mass difference \( \Delta M = M_1-M_{Z^0}\). 
Rearranging Eq.~\ref{eqn_tanb} 
the mass of the heavier gauge boson takes the form
\be
M_2^2 = \frac{(1+\tan^2\theta)(M_1-\Delta M)^2-M_1^2}{\tan^2\theta}.
\ee
Thus, in our study we consider $\theta$ and $\Delta M$ as the two independent parameters
of the mixing. The importance of mixing in $W$-pair production at ILC is clearly visible
with the fact that the
new gauge sector does not interact with the SM gauge sector directly, and therefore
the $W$ boson does not couple directly to $Z'$ at tree level. 
Its couplings to the mass eigenstates $Z_2$ arises through mixing, along with a corresponding
weakening in its coupling to the lighter mass eigenstate, $Z_1$. These couplings are 
given by
\be
g_{WWZ_1} = g_{WWZ^0}~\cos\theta,~~~~~~~~~~~~
g_{WWZ_2} = g_{WWZ^0}~\sin\theta,
\ee
where $g_{WWZ^0}$ is the SM coupling. Experimental constraints on the mixing 
limits the value of $\theta$ to be not larger than a few times $10^{-3}$. 
At the same time, when the Higgs structure of the model is known, one may compute the mixing angle. 
In such a case, mixing angle can be expressed as
\be
\theta = C~\sqrt{\frac{5}{3}\lambda}\sin\theta_W~\frac{M_{Z_1}^2}{M_{Z_2}^2},
\ee
where $\lambda$ is a parameter of order unity, $C$ is function of the 
VEV's of the Higgs fields and their $U(1)'$ charges and $\theta_W$ is the usual 
Weinberg mixing angle. In Table~\ref{table_Ctheta}
we present an illustrative case of $E_6$ models as discussed in 
Ref.\cite{Erler:2009jh}. 
The table also gives the mixing angle, $\theta$ and $\Delta M$ in each case 
corresponding to a representative value of $M_{Z_2}=1$ TeV.  

\begin{table}[ht]
\begin{center}
\begin{tabular}{|l|l|c|c|}
\hline
&&&\\
Model&$C$ (range)&$\theta$&$\Delta M$ (MeV)\\[2mm] \cline{1-4}
&&&\\
$E_6(\chi)$& $\left[-\frac{3}{\sqrt{10}},\frac{2}{\sqrt{10}}\right]$&$\left[-.0037,.0025\right]$&$\left[75,33\right]$\\[2mm]
$E_6(\psi)$& $\left[-\sqrt{\frac{2}{3}},+\sqrt{\frac{2}{3}}\right]$&$\left[-.0032,.0032\right]$&$56$\\[2mm]
$E_6(\eta)$& $\left[-\frac{1}{\sqrt{15}},+\frac{4}{\sqrt{15}}\right]$&$\left[-.0010,.0040\right]$&$\left[6,89\right]$\\[2mm]
$LRSM$&$\left[-\frac{1}{\alpha_{LR}}\sqrt{\frac{3}{5}},+\alpha_{LR}\sqrt{\frac{3}{5}}\right]$&$\left[-.0019,.0048\right]$&$\left[20,124\right]$\\[2mm]\cline{1-4}
\end{tabular}
\caption{Mixing angle ($\theta$) and $\Delta M$ corresponding to a $Z_2$ of mass 1 TeV in
different models considered. The parameter in the left-right symmetric model (LRSM) takes a value,
$\alpha_{LR}=\sqrt{1-2\sin^2\theta_W}/\sin\theta_W$.  }
\label{table_Ctheta}
\end{center}
\end{table}

We now turn our attention to the $Zee$ coupling. 
The fermion couplings are highly model 
dependent. In the following we will very briefly 
describe this coupling in the models considered. A detailed 
analysis can be found in the literature including \cite{Langacker}. 
The neutral current interactions of electrons with the gauge bosons
are given by the Lagrangian term,
\be
{\cal L}_{NC}=J^\mu_{em}A_\mu+J^\mu_{SM}Z^0_\mu+J'^\mu Z'_\mu,
\ee 
where $J^\mu_{em}$ is the electromagnetic current with which 
the photon interact, 
$J^\mu_{SM}$ is the current with which the SM neutral gauge boson, 
$Z^0$ interact, and 
$J'^\mu$ is the current with which the new neutral gauge boson, $Z'$ interact. 
In terms of the projection operators $P_{L,R}=(1\mp \gamma^5)/2$ the currents take the form
\be
J_i^\mu=-\bar\psi_e\gamma^\mu(g^e_{iL}P_L+g^e_{iR}P_R)\psi_e. 
\label{eqn_currents}
\ee
The SM couplings involved in $J^\mu_{SM}$ are expressed 
in terms of the Weinberg mixing angle, $\theta_W$ as
\be 
g^f_L=\frac{e}{\sin\theta_W\cos\theta_W}(-\frac{1}{2}+~\sin^2\theta_W),~~~~~~~
g^f_R=\frac{e\sin\theta_W}{\cos\theta_W}.
\ee
Fermion couplings corresponding to the current, $J'^\mu$ in the cases of models considered in
our study are tabulated in Table~\ref{table_fc}.
\begin{table}[ht]
\begin{center}
\begin{tabular}{|l|c|c|}
\hline
&&\\
Model&$g'^e_L$& $g'^e_R$ \\[2mm] \cline{1-3}
&&\\
$E_6(\chi)$&$\sqrt{\frac{3}{8}}\frac{e}{\cos\theta_W}$&$\sqrt{\frac{1}{24}}\frac{e}{\cos\theta_W}$\\[2mm]
$E_6(\psi)$&$\frac{\sqrt{10}e}{12\cos\theta_W}$& $-\frac{\sqrt{10}e}{12\cos\theta_W}$\\[2mm]
$E_6(\eta)$&$\frac{e}{6\cos\theta_W}$& $\frac{e}{3\cos\theta_W}$\\[2mm] 
$LRSM$&$\frac{e}{\cos\theta_W}\frac{1}{2\alpha_{LR}}$
&$\frac{e}{\cos\theta_W}(\frac{1}{2\alpha_{LR}}-\frac{\alpha_{LR}}{2})$\\[2mm]
$ALRSM$&$\frac{e}{\cos\theta_W \sin^2\theta_W}
\frac{1}{\alpha_{LR}}
(-\frac{1}{2}+\sin^2\theta_W)$&$\frac{e}{\cos\theta_W \sin^2\theta_W}
\frac{1}{\alpha_{LR}}
(-\frac{1}{2}+\frac{3}{2}\sin^2\theta_W)$\\[2mm]\cline{1-3}
\end{tabular}
\caption{$Z'ee$ couplings (Eq. \ref{eqn_currents}) 
in different models considered. The parameter $\alpha_{LR}=\sqrt{1-2\sin^2\theta_W}/\sin\theta_W$.}
\label{table_fc}
\end{center}
\end{table}

In the next section, we will compare the effects found in the above
models, with those arising in the Littlest Higgs scenario
whose collider signature at the ILC with polarized beams
was recently considered in ref.~\cite{APP}.
Briefly stated,
in this scenario, the Higgs fields are considered to be 
the Nambu-Goldstone Bosons
(NGB) of the non-linear realization of some global symmetry breaking. 

 In our numerical analysis for LHM, there are two free parameters
$f$ and $\theta_H$. As argued by \cite{Dobado}, precision electroweak 
measurements restrict the parameters to be $f\sim 1$ TeV and 
$0.1<\cos\theta_H<0.9$. 
In our study we consider a 
value of $f = 1$ TeV and  $\cos\theta_H = 0.3$ satisfying these restrictions.

\section{Analyses of $e^+e^-\rightarrow WW$} \label{Analysis}
In this section we present the results of our numerical analysis to probe the 
$Z'$ models through the process $e^+e^-\rightarrow WW$ at the ILC. This process gets an additional
$s$-channel contribution through the exchange of the heavier gauge bosons, $Z_2$. 
Only the SM component of $Z_2$ interacts with $W$, and therefore this additional $s$-channel 
contribution is not present in the absence of mixing. At the same time, $Z_1$ becomes the SM 
gauge boson, $Z^0$ with the standard couplings, when there is no mixing.
The $t$-channel with $\nu$-exchange is not affected by new physics here 
as $W$ interactions with SM fermions proceed through standard couplings. Thus the $e^+e^-\rightarrow WW$ process directly probes
the $Z^0-Z'$ mixing, unlike processes like $e^+e^-\rightarrow f\bar f$, which gets contribution from
the presence of $Z'$ even when there is no mixing.

\begin{figure}[htb]
\begin{center}
\includegraphics[width=10 cm,height=25mm]{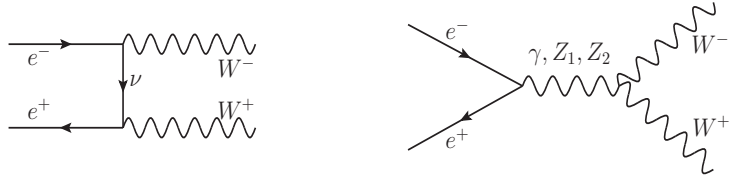}
\caption{Feynman diagrams contributing to the process $e^+e^-\rightarrow
W^+W^-$ in the $Z'$ models. }
\label{fig:FD}
\end{center}
\end{figure}

Thus for the process $e^+e^-\rightarrow W^+W^-$, 
we have an $s$-channel process with the
exchange of the heavy neutral gauge boson, $Z_2$, in addition to the standard 
channels
as shown in Fig.~\ref{fig:FD}.

The three-point gauge couplings involving $WW$ are given by:
\[
V^\mu (k_1)W^\nu (k_2)W^\rho (k_3)=ig_{VWW}~\left[
g^{\mu\nu}(k_1-k_2)^\rho+
g^{\nu\rho}(k_2-k_3)^\mu+
g^{\rho\mu}(k_3-k_1)^\nu \right],
\]
where all the momenta are considered outflowing, and $V\equiv \gamma,~Z_1,~Z_2$.

Direct search results from $p\bar p$ collisions at TeVatron as well as precision electroweak
analysis constrain the parameters discussed in the previous section \cite{RPP}. In most cases $Z_2$ mass 
slightly smaller than 1 TeV is permitted. This is translated into allowed range of couplings in a specific
version of $E_6$ model as presented in Table~\ref{table_Ctheta}. In our numerical studies we consider a conservative
value of 0.003 and 100 MeV for the mixing angle and the parameter $\Delta M$, respectively. The above parameter choices
correspond to a $Z_2$ of mass $\sim 1400$ GeV .

For a non-exhaustive list of phenomenological studies of $Z'$ in the context of LHC as well as ILC, see reference in
\cite{Osland1, Osland2, Pheno1, Pheno2}.
Previous phenomenological studies of $e^+e^-\rightarrow WW$ process in the context of $Z'$ have considered different 
observables at the production level \cite{PP,Pheno2}, and have obtained the reach of ILC in probing the model. In this work 
our main focus is on various observables constructed with the decay products of the $W$'s produced. One obvious 
advantage is that these observables do not require full reconstruction of the $WW$ events, unlike in the former case.
In this section, we will first report our analysis of the process at the production level with a study of the total 
cross section, and subsequently analyze different decay distributions and other observables.

\subsection{The total cross section}
We compute the total cross section incorporating beam polarization 
using the helicity amplitudes given in ref.~\cite{Hagiwara} with the new 
couplings and with the added contribution due to the exchange of $Z_{2}$.
With beam polarization, in general,
the polarized cross section may be expressed
as:

\begin{eqnarray}
\sigma(e^{+}e^{-}\rightarrow W^{+}W^{-}) &=&\frac{1}{4}\left[(1+P_{e^{-}}).(1-P_{e^{+}})\sigma^{RL}\right. \nonumber \\
    &&+\left.(1-P_{e^{-}}).(1+P_{e^{+}})\sigma^{LR}\right],
\end{eqnarray}
where \(~~
\sigma^{RL}=\sigma(e^{+}_Le^{-}_R\rightarrow W^{+}W^{-}) ~~{\rm and}~~
\sigma^{LR}=\sigma(e^{+}_Re^{-}_L\rightarrow W^{+}W^{-}), 
\) with
$e_{L,R}$ representing the left- and right-polarized electrons 
(and positrons), respectively. The degree of polarization is defined 
as: 
$P_{e}=(N_R-N_L)/(N_R+N_L)$, where $N_{L,R}$ denote the number 
of left-polarized and right-polarized electrons (and positrons), respectively.
More than 80\% of electron beam polarization and large positron beam 
polarization are expected to be achieved at ILC. In our analysis we consider
both the ideal possibility of 100\% polarization along with the realistic
polarization of the beams that will be achieved at ILC.

\begin{figure}[htb]
\hspace{2.8cm}
\includegraphics[width=6 cm,height=6 cm]{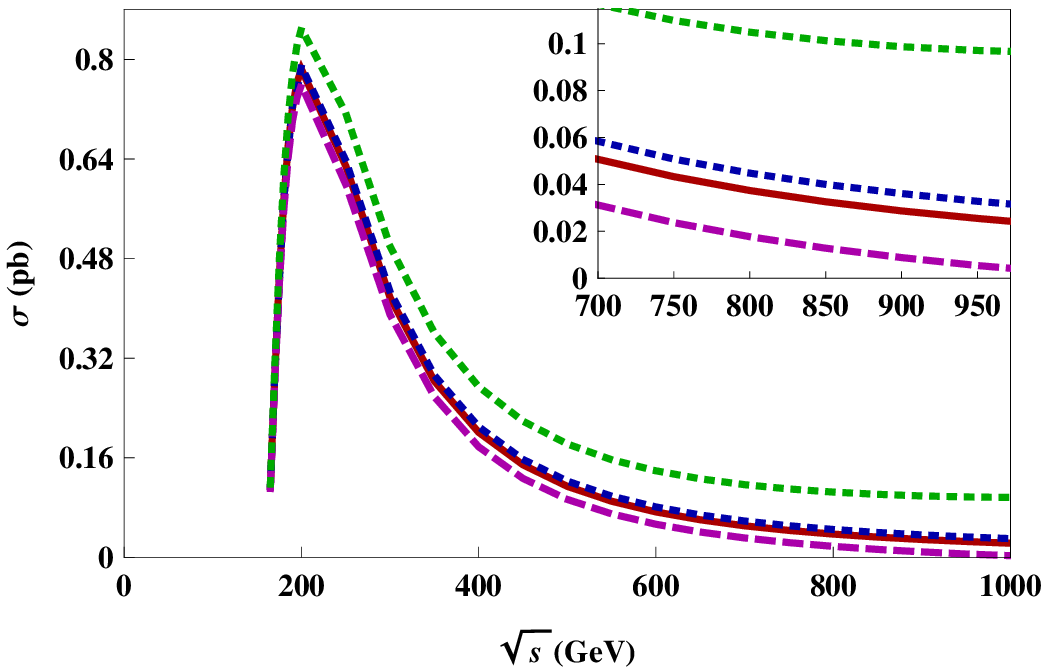}
\caption{Total cross section for $W^{+}$ $W^{-}$ production in an $e^{+}e^{-}$ collision for SM[Red-Solid], 
ALRSM, LRSM [Magenta-Dashed], LHM [Green-Dotted]
and $E_6(\chi)$ [Blue-Dotted] with polarized beams 
with $P_{e^{-}}$=1 and $P_{e^{+}}$=-1. For $U(1)'$ models, $\theta=0.003$ and 
$\Delta M=100$ MeV are considered. The parameter values of $f$=1 TeV and $c$=0.3 are used for LHM}
\label{fig_sig}
\end{figure}

The models considered here are insensitive to the total cross section
even at higher centre of mass energies.  The inclusion of beam polarizations 
has no significant effect.  In Table~\ref{tab_cs}, we present the total production
cross section for SM and $Z'$ models considering both unpolarized and polarized beams.
It can be seen that at $\sqrt{s}$=500 GeV, the deviation 
for different $Z'$ models from the SM using unpolarized beams is about 0.05 - 0.3\%. 
The deviation is slightly
increased to about 0.7\% at $\sqrt{s}$=800 GeV.  Switching on the polarization 
with left-handed electrons and right-handed positrons 
it is seen that with this choice of polarization the cross section is about
four times more than the unpolarized case. Although the deviation in this
case follows the same pattern as the unpolarized one, but due to the larger
size of the cross section $Z - Z'$ mixing can be studied more effectively.
However with right-handed electrons and left-handed positrons, due to the absence
of the dominant t channel the mixing effect is more pronounced, even though the 
cross section is very small compared to the other two cases.  This particular
combination thus leads to an increased signal by background ratio.  It can be seen that the deviation
is about 24\% for different $E_6$ models and 50\% for the LRSM and ALRSM model at $\sqrt{s}$=800 GeV,
with 100\% beam polarization.
We have only presented the figure for this particular case. In Fig.~\ref{fig_sig} we present
the total production cross-section 
in the case of SM and $E_6(\chi)$ as this has a maximum deviation compared
to other $E_6$ models. Moreover, it can be seen that the percent deviation
of LRSM and ALRSM are equal with this specific choice
of beam polarization.  LHM is also plotted for comparison.  Since the gauge structure
of LHM is different from the other $Z'$ models considered here, it behaves differently 
from them.  It is not much constrained compared to the other models as can be seen from
Fig.~\ref{fig_sig}. Taking into account the polarization which will be achieved
at ILC, different $E_6$ models show about 4\% deviation from the SM. The percent 
deviation for LRSM and ALRSM is about 9\% compared to 50\% obtained in the ideal case.

\begin{table}
%\begin{center}
%\fontsize{26}{17}\selectfont Huge text
\begin{tabular}{|c|c||c||c|c|c|c|} \hline
       $P_e^-$	 &$P_e^+$     &model              &$\sqrt{s}=$300 GeV  &$\sqrt{s}=$500 GeV  &$\sqrt{s}=$800 GeV   &$\sqrt{s}=$1000 GeV  \\[1mm]  \hline
                   &            &SM                 &13.598   & 7.208   & 3.744    & 2.692      \\ [1mm]
                &            &$E_6(\chi)$           &13.618   &7.227    &3.767   & 2.724        \\[1mm]
                &            & $E_6(\psi)$          &13.602   & 7.212  &3.749    &2.699          \\[1mm]
       0        &0           & $E_6(\eta)$          & 13.612   & 7.223   & 3.762    & 2.717   \\[1mm] 
                &            &LRSM                & 13.601   &7.212  &3.750      & 2.701       \\[1mm]
                &            &ALRSM                & 13.571  &7.183   &3.720      &2.666           \\[1mm]
                 &            &LHM            & 13.679  &7.347   &3.867      &2.973           \\[1mm]  \cline{1-7} 
                &            &SM                  &53.973   &28.716   &14.940      & 10.746       \\[1mm] 
                &            &$E_6(\chi)$         &54.042   &28.777   &14.994      &10.798         \\[1mm]
         &     &$E_6(\psi)$        &54.005   &28.743  &14.963     &10.767          \\[1mm]
-1   &1        &$E_6(\eta)$    &54.038   &28.774     &14.991      &10.795       \\[1mm]
               &        &LRSM                  & 54.013  &28.753   &14.981     &10.800        \\[1mm] 
                &            &ALRSM                &53.895   &28.640     &14.862    &10.660       \\[1mm]
                 &            &LHM      &54.214   &29.205     &15.362    &11.796       \\[1mm]        \cline{1-7} 
               &            &SM                 &41.023 &21.826   &11.355     &8.167  \\[1mm] 
                &            &$E_6(\chi)$        & 41.076 &21.871  &11.396    & 8.206         \\[1mm]
                &            &$E_6(\psi)$        & 41.048  & 21.845 & 11.372  & 8.183         \\[1mm]
       -0.9        &0.6          &$E_6(\eta)$        & 41.073  & 21.869   & 11.393 & 8.204         \\ [1mm]
                &            &LRSM                 & 41.053 & 21.850 & 11.376  & 8.187         \\[1mm]
                &            &ALRSM                & 40.964 & 21.767 & 11.295  & 8.102     \\ [1mm]
                &            &LHM                & 41.208 & 22.198 & 11.676  & 8.966     \\       \hline   
               &            &SM                 &0.418  &0.114   &0.037     &0.023   \\[1mm] 
                &            &$E_6(\chi)$        &0.427  &0.122   &0.045    &0.030         \\[1mm]
                &            &$E_6(\psi)$        &0.401   &0.103  &0.028   &0.015          \\[1mm]
       1        &-1          &$E_6(\eta)$        & 0.411   &0.110    &0.034  &0.020          \\ [1mm]
                &            &LRSM                 &0.390  &0.093   &0.018   &0.003          \\[1mm]
                &            &ALRSM                &0.390  &0.093  &0.018   &0.003      \\ [1mm]
                &            &LHM                &0.500  &0.182  &0.105   &0.096      \\       \hline   
               &            &SM                 & 0.857 & 0.374  & 0.178    & 0.125  \\[1mm] 
                &            &$E_6(\chi)$        & 0.865  & 0.381 &  0.184 & 0.131        \\[1mm]
                &            &$E_6(\psi)$        & 0.845  & 0.365 & 0.171  & 0.119         \\[1mm]
       0.9        &-0.6          &$E_6(\eta)$        & 0.853   & 0.371   & 0.176 & 0.123         \\ [1mm]
                &            &LRSM                 & 0.837 & 0.360  & 0.167  & 0.115         \\[1mm]
                &            &ALRSM                & 0.835 & 0.357 & 0.162  & 0.109     \\ [1mm]
                &            &LHM                & 0.922 & 0.431 & 0.233  & 0.191 \\       \hline   

\end{tabular}
%\end{center}
\caption{Total cross section (in pb) for different models
with both polarized and unpolarized beams. The parameter used for
LHM is $f$=1 TeV and $c$=0.3, and for the $Z'$ models $\theta=0.003$ and
$\Delta M=100$ MeV}
\label{tab_cs}
\end{table} 

\subsection{Double Energy Distribution}
In order to exploit further the process at hand, it is profitable
to consider the decays of one or both the $W$'s.  
Let us consider $e^{+}e^{-}\rightarrow W^{+}W^{-}$ with 
both $W$'s decaying into leptons. The differential cross section in this case
can be expressed as
\begin{eqnarray}
& \displaystyle
\frac{d\sigma}{d\cos\Theta~d\cos\theta^*_-~d\phi^*_-d\cos\theta^*_+~d\phi^*_+} =
& \nonumber \\
& \displaystyle 
\frac{9\beta}{8192\pi^3s}B(W^-\rightarrow l^-\bar\nu_l)B(W^+\rightarrow l^+\nu_l)
{\cal P}^{\lambda \bar{\lambda}}_{\lambda' \bar{\lambda}'}{\cal D}^{\lambda}_{\lambda'}
~\bar{{\cal D}}^{\bar{\lambda}}_{\bar{\lambda}'} ,
\label{eqn_dsig_hagi}
\end{eqnarray}
where $\Theta$ is the scattering angle and $\beta = \sqrt{(1-4m_W^2/s)}$ is the velocity of the $W$ 
in the centre of mass frame, $\sqrt{s}$ being the centre of mass energy. The other angles,
$\theta^*_\mp$ and $\phi^*_\mp$ are the polar and azimuthal angles of the 
lepton/antilepton in the rest-mass frame of $W^\mp$ with boost direction of $W^-$ along the
$z$-axis, respectively. The production and decay tensors, 
${\cal P}^{\lambda \bar{\lambda}}_{\lambda' \bar{\lambda}'}$, 
${\cal D}^{\lambda}_{\lambda'}$ and $\bar{{\cal D}}^{\bar{\lambda}}_{\bar{\lambda}'}$ are given in \cite{Hagiwara}. 
The energy of the $l^\mp$ in the centre of mass frame is related to $\theta^*_\mp$ in the following way:
\be
E_{l^\mp}=\frac{\sqrt{s}}{4}~\left(1\pm \beta\cos\theta^*_{\mp}\right),
\ee
%where $\gamma=1/\sqrt{(1-\beta^2)}$.
This allows us to obtain the double energy differential cross section from Eq.~\ref{eqn_dsig_hagi} as

\be
\frac{d\sigma}{dE_{l^-}dE_{l^+}} =\int \left(\frac{2}{\beta\gamma m_W}\right)^2~
\frac{d\sigma}{d\cos\Theta~d\cos\theta^*_-~d\phi^*_-d\cos\theta^*_+~d\phi^*_+}~d\cos\Theta~d\phi^*_-~d\phi^*_+. 
\label{eqn_EE}
\ee
Following the notation of Dicus and Kallianpur \cite{DK},
the energies of the leptons $E_{l^\pm}$ are expressed as dimensionless variables
$X_{l^\pm}$ defined as:

\begin{equation}
X_{l^\pm} = \frac{2}{\beta\sqrt{s}}\left(E_{l^\pm}-\frac{\sqrt{s}}{4}(1-\beta)\right).
\end{equation}
$X_{l^\pm}$ varies between 0 and 1. 
In Table~\ref{table_ded} we present the percentage of events in bins of $X_{l^-}$ and $X_{l^+}$ obtained from
\begin{eqnarray*}
\frac{1}{\sigma}  \frac {d^2\sigma}{dX_{l^+}dX_{l^-}} 
\end{eqnarray*}
at $\sqrt{s}=800$ GeV using unpolarized beams. As expected in the case of SM the distribution peaks at
maximum values of $X_{l^-}$ and $X_{l^+}$. This behaviour dominates in other models as well. 
Since the matrix in Table~\ref{table_ded} is symmetric under interchange
of $X_{l^-}$ and $X_{l^+}$ (see Ref.~\cite{DK} for explicit expression), we have shown only the upper half. 
We have numerically checked that the matrix is indeed symmetric.
\begin{table}
\begin{center}
%\fontsize{26}{17}\selectfont Huge text
\begin{tabular}{|c c| c c c c c c c c c c} \hline
   &$X_{l^+}$  & &0.2 & &0.4 & &0.6 & &0.8 & &1.0          \\ 
$X_{l^-}$  & & & & & & & & & & &  \\   \hline
1.0    & & & & & & & & & & &   \\    
      &  &0.428 & &2.705 &  &7.174 & &13.837 & & $22.693^{\tiny a}$ \\
       &  &0.433  &  &2.707 & &7.174  & &13.773 & &$22.566^{\tiny b}$ \\
         &  & 0.429 & &2.705 &  &7.170  &  &13.824 &  &$22.665^{\tiny c}$  \\
         &  &0.431  & &2.707  & &7.159  &  &13.788 & &$22.594^{\tiny d}$ \\
         &  &0.429  & &2.706  & &7.170  & &13.823 & &$22.664^{\tiny e}$ \\
         &  &0.422  &  &2.702  & &7.197 & &13.907 & &$22.832^{\tiny f}$ \\
         &  &0.465  &  &2.716  & &7.023 & &13.386 & &$21.804^{\tiny g}$ \\
0.8       &     &  &  & & & & & & & \\ 
        &  &0.355 & &1.747 & &4.459 & &8.488 & & \\ 
        &  &0.368 & &1.772 & &4.474 & &8.475 & &   \\
        &  &0.358 & &1.753 & &4.462 & &8.486 & &  \\
        &  &0.365 & &1.767 & &4.471 & &8.478 & &   \\
        &  &0.358 & &1.753 & &4.462 & &8.486 & &  \\
        &  &0.341 &  &1.720 &  &4.441 &  &8.503 &  & \\
        &  &0.450 & &1.920 &  &4.567 &  &8.389   &  & \\
0.6    & & & & & & & & & & \\
      & &0.364 & &1.053 & &2.418 & & & & \\
      & &0.379 & &1.087 & &2.452 & & & &  \\
      & &0.367 & &1.060 & &2.425 & & & &  \\
      & &0.376 & &1.080 & &2.445 & & & &  \\
      & &0.367 &  &1.060 & &2.426 & & & & \\
      & &0.348 & &1.016 & &2.381 & & & &  \\
      & &0.475 & &1.293 & &2.657 & & & & &  \\
0.4   & & & & & & & & & & \\
      & &0.454 & &0.622 & & & & & & \\
      & &0.466 & &0.652 & & & & & &  \\
      & &0.456 & &0.628 & & & & & & \\
      & &0.463 & &0.645 & & & & & &  \\
      & &0.456 & &0.629 & & & & & & \\
      & &0.442 & &0.590 & & & & & & \\
      & &0.541 & &0.833 & & & & & & \\
0.2   & & & & & & & & & &  \\
      & &0.626 & & & & & & & & \\ 
      & &0.628 & & & & & & & & \\ 
      & &0.626  & & & & & & & & \\
      & &0.628 & & & & & & & & \\ 
      & &0.626 & & & & & & & & \\
      & &0.623 & & & & & & & & \\
      & &0.646 & & & & & & & & \\ \hline    
\end{tabular}
\end{center}
\caption{Percentage of events at $\sqrt{s}$=800 GeV with unpolarized beams in
bins of $X_{l^-}$ and $X_{l^+}$ corresponding to different models: 
$a$ =SM, $b$=$E_6(\chi)$, $c$=$E_6(\psi)$, $d$=$E_6(\eta)$,  $e$=LRSM, $f$=ALRSM and
$g$=LHM. The parameters for $U(1)'$ models, $\theta$=0.003 
and $\delta M$=100 MeV, while for LHM $f=1$ TeV and $c=0.3$ are used}
\label{table_ded}
\end{table} 

A combination of the effects result in about 5\% deviation in the case of 
$E_6(\chi)$ and ALRSM models, but with a qualitative difference. In the case of
$E_6(\chi)$ it is an enhancement, whereas in the ALRSM case there is a reduction. The effect
is reduced to smaller than 4\% in the case of $E_6(\eta)$ model, while in the 
case of $E_6(\psi)$ and LRSM models it is a very negligible contribution of about 1\%. The most 
sensitive bin in all the cases is the one with $0.2\leq (X_{l^-},X_{l^+})\leq 0.4$, with 
about 0.6\% of the total events in the case of SM. Such small deviation of a few percent is therefore
hard to detect even at a high luminosity machine like the ILC. On the other hand, the LHM model, while
showing similar qualitative behaviour, deviates from the SM value by about 34\% in the same bin, leaving 
scope of detection at the ILC. 

\begin{figure}[htb]
\hspace{2.8cm}
\includegraphics[width=5 cm,height=5 cm]{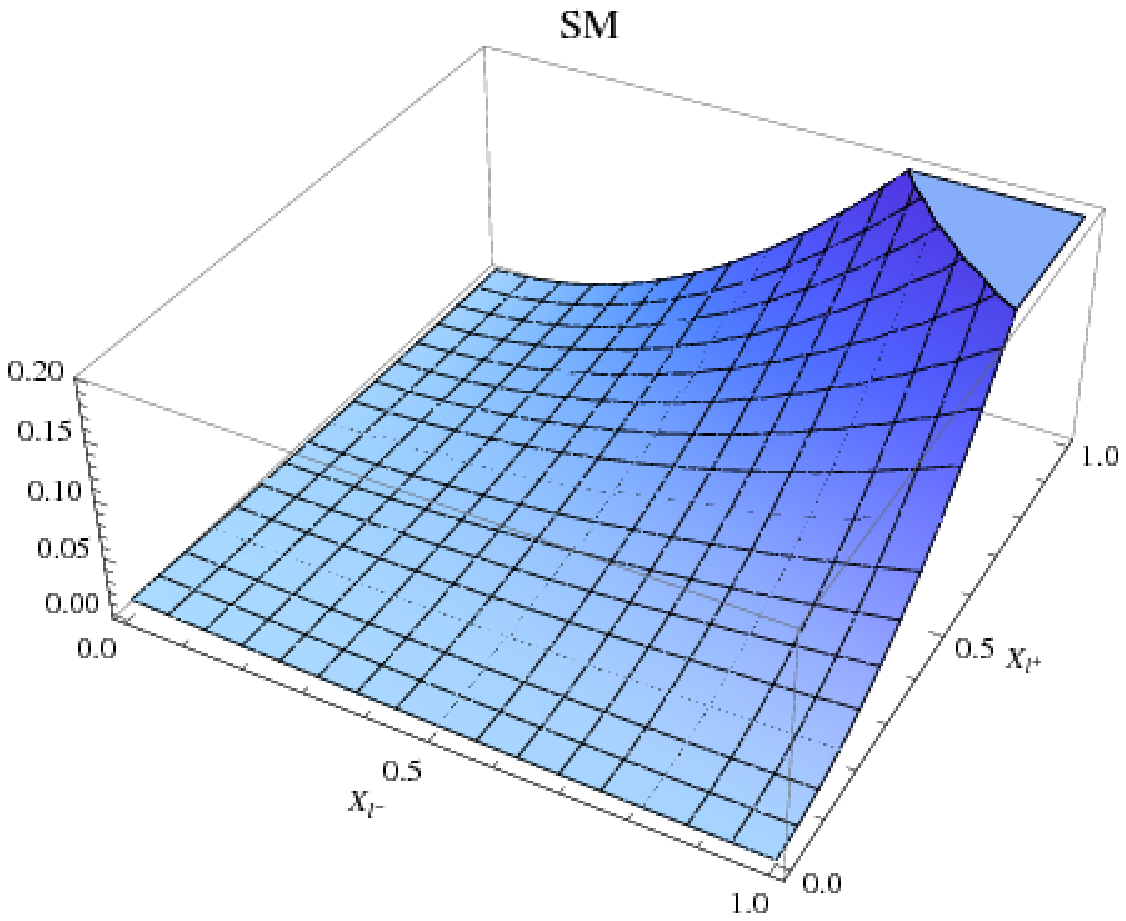}
\caption{Double energy distribution 
for the SM at $\sqrt{s}=800$ GeV with
unpolarized beams}
\label{fig_sigSMunpol}
\end{figure}

\begin{figure}[b!]
\includegraphics[width=5 cm,height=5 cm]{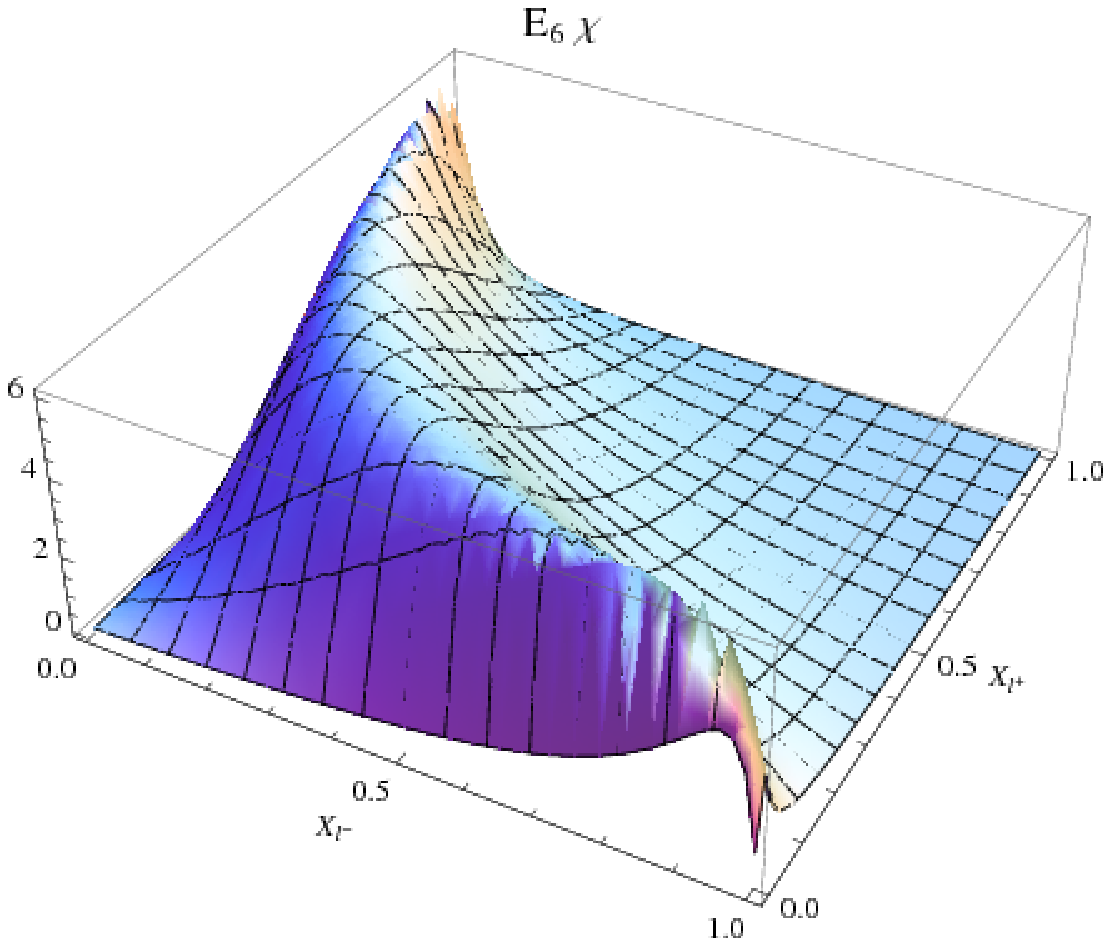}
\hspace{0.2cm}
\includegraphics[width=5 cm,height=5 cm]{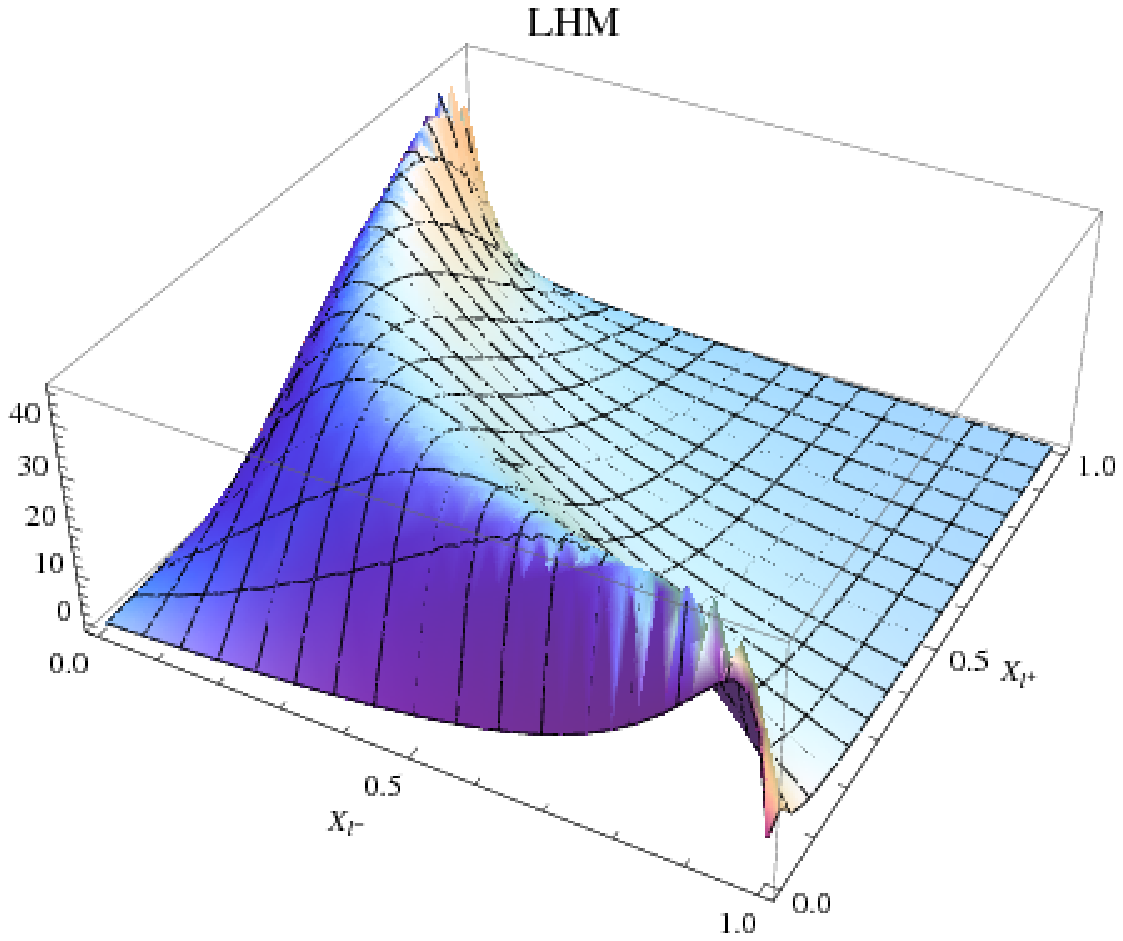}
\caption{Double energy distribution showing the percentage deviation
for the two models $(a)$ $E_6(\chi)$ and $(b)$ LHM from the SM at $\sqrt{s}=800$ GeV with
unpolarized beams. In the case of $E_6(\chi)$ $\theta=0.003$ and $\Delta M=100$ MeV, 
and in the case of LHM $f=1$ TeV and $c=0.3$ are used}
\label{fig_sigEEunpol}
\end{figure}

In the above analysis we have not made any attempt to optimize our results by considering different 
binning options. We expect the qualitative behaviour to remain more or less the same even in the optimal case.
This is supported by Fig.~\ref{fig_sigSMunpol} and
Fig.~\ref{fig_sigEEunpol}, where we plot the double energy distribution for SM and the 
deviation from SM expressed as
 \begin{eqnarray*}
\frac{\left.\frac{d^2\sigma}{dX_{l^-}dX_{l^+}}\right|_{model}-
\left.\frac{d^2\sigma}{dX_{l^-}dX_{l^+}}\right|_{SM}}{\left.\frac{d^2\sigma}{dX_{l^-}dX_{l^+}}\right|_{SM}}.
\end{eqnarray*}

We can see from Fig.~\ref{fig_sigSMunpol} as discussed before, in the case of SM the distribution peaks at
maximum values of $X_{l^-}$ and $X_{l^+}$. 

The use of right-handed electron
beam and left-handed positron beam is expected to have a much larger sensitivity to all the models. In the case of
LHM, while the new gauge boson do not contribute, owing to the fact that this does not couple to the right-handed 
electrons, the changed fermionic couplings of the SM gauge boson provides substantial effect \cite{APP}. In this case,
the $t$-channel contribution is not present, and therefore both the SM as well as new physics
contributions show a symmetric behaviour. The difference in the two cases is, as expected, a constant shift, either
positive or negative.  
At $\sqrt{s}=800$ GeV, about 50\% deviation is seen for both LRSM and ALRSM models, while $E_6(\psi)$ has about 40\% deviation.
The other two models show slightly reduced sensitivity with about 30\% and 12\% deviations in the case of $E_6(\chi)$ and $E_6(\eta)$
models respectively. 

\subsection{Single Energy Distribution}

\begin{figure}[ht]
\begin{minipage}[b]{0.5\linewidth}
\centering
%\vspace{-0.4cm}
\includegraphics[scale=0.6]{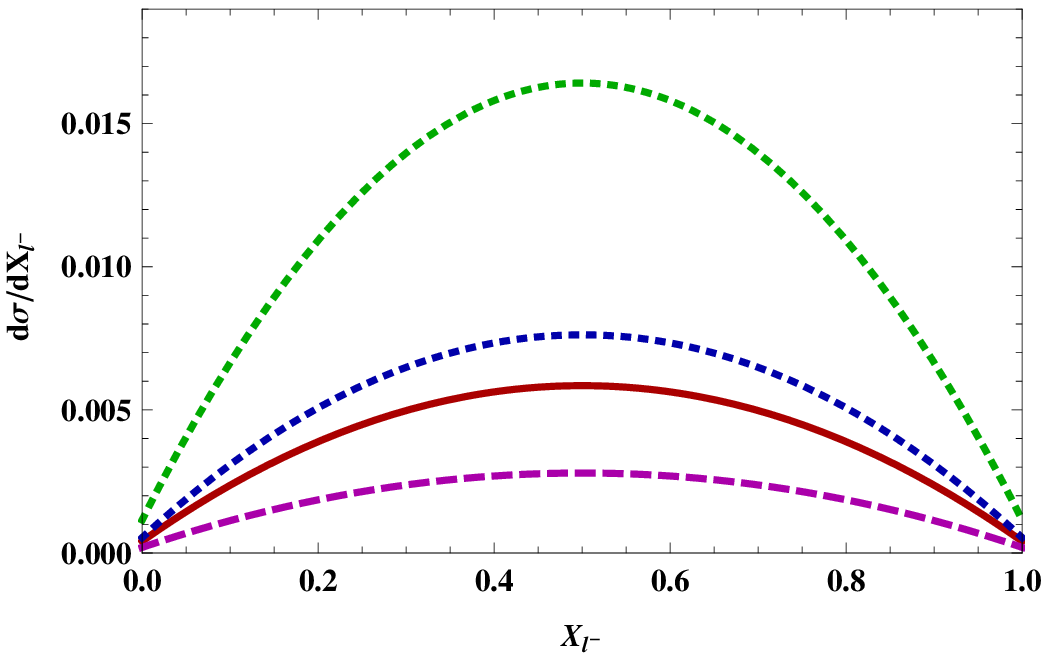}
\caption{{\small Energy distribution of the secondary leptons.
with polarized beams with $P_{e^{-}}$=1 and $P_{e^{+}}$=-1 at $\sqrt{s}=800$ GeV. 
The parameters used are $\Delta M$=100 MeV 
and $\theta$=0.003 for $U(1)'$ type of models, and $f=1$ TeV and $c=0.3$ in the case of LHM.  Different curves correspond to
 SM (Red-Solid), LHM (Green-Dotted), $E_6(\chi)$ (Blue-Dotted), and ALRSM and LRSM (Magenta-Dashed).}}
\label{fig_E}
\end{minipage}
\hspace{0.5cm}
\begin{minipage}[b]{0.5\linewidth}
\centering
\includegraphics[scale=0.6]{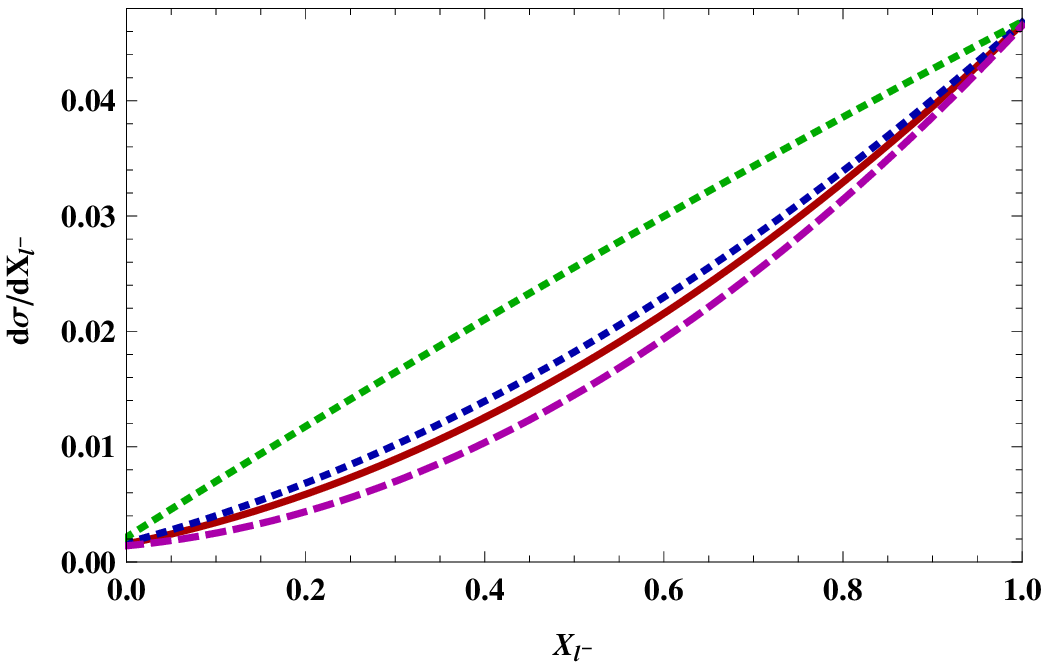}
\caption{{\small Energy distribution of the secondary leptons.
with polarized beams with $P_{e^{-}}$=0.9 and $P_{e^{+}}$=-0.6 at $\sqrt{s}=800$ GeV. 
The parameters used are $\Delta M$=100 MeV 
and $\theta$=0.003 for $U(1)'$ type of models, and $f=1$ TeV and $c=0.3$ in the case of LHM.  Different curves correspond to
 SM (Red-Solid), LHM (Green-Dotted), $E_6(\chi)$ (Blue-Dotted), and ALRSM and LRSM (Magenta-Dashed).}}
\label{fig_ER}
\end{minipage}
\end{figure}

Energy distribution of the secondary lepton obtained by
integrating the double energy distribution in Eqn.~\ref{eqn_EE} over $E_{l^+}$, is another observable that might give
a handle on the new effects. In Figs.~\ref{fig_E},~\ref{fig_ER}
we plot the energy distribution of the lepton for the case of 
$\sqrt{s}=800$ GeV for both ideal
and realistic degrees of beam polarizations. 
The signs of the beam polarization are chosen so as to essentially
switch off the SM t-channel contribution, so as to enhance the
effects of the new physics.
There is no appreciable deviation 
in the case of unpolarized beams 
(and in the case of left-handed electron beams). 
We notice that the LRSM and ALRSM cases behave qualitatively differently 
compared to the $E_6$ models and the LHM. While in the former case there is a reducing effect for the entire range of $X_{l^-}$, the 
latter has an increasing effect. This could be used as a discriminating factor between the left-right symmetric models and the other models used.

Note that in the case of realistic degrees of polarization since the
neutrino t- channel effects are present, the panels comparing
ideal and realistic degrees of freedom appear quite different
as functions of $X_{l^-}$.

\subsection{Angular Spectrum of the Secondary Lepton}

\begin{figure}[ht]
\begin{minipage}[ht]{0.5\linewidth}
\centering
%\vspace{-0.4cm}
\includegraphics[scale=0.6]{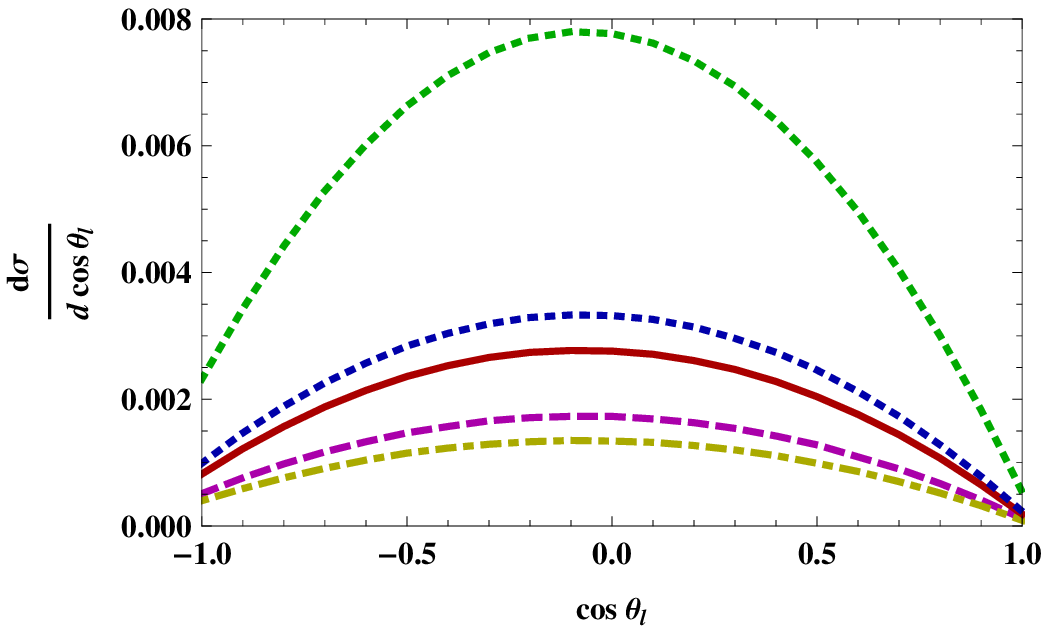}
\caption{{\small Angular distribution of one of the secondary leptons at $\sqrt{s}$=800 GeV using 
polarized beams with $P_{e^{-}}$=1 and $P_{e^{+}}$= -1 in the case of  
SM (Red-Solid), LHM (Green-Dotted), $E_6(\chi)$ (Blue-Dotted), ALRSM (Yellow-DotDashed)
and LRSM (Magenta-Dashed). 
The parameters used are $\Delta M$=100 MeV and $\theta=0.003$ in the case of $E_6$ and LR models, and
$f=1$ TeV and $c=0.3$ in the case of LHM.}}
\label{fig_angdist}
\end{minipage}
\hspace{0.5cm}
\begin{minipage}[ht]{0.5\linewidth}
\centering
\includegraphics[scale=0.6]{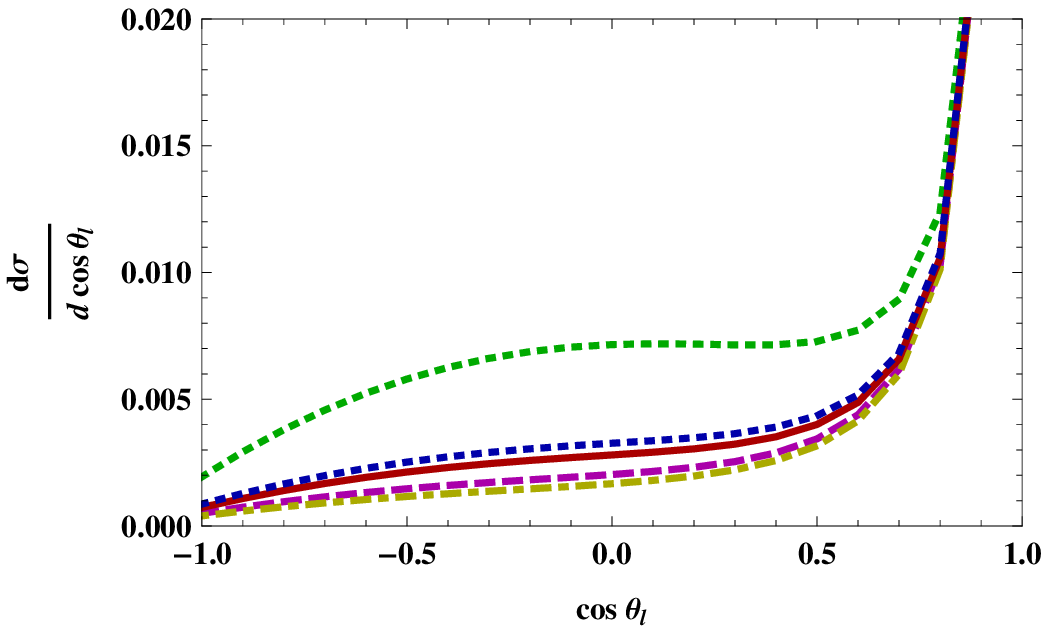}
\caption{{\small Angular distribution of one of the secondary leptons at $\sqrt{s}$=800 GeV using 
polarized beams with $P_{e^{-}}$=0.9 and $P_{e^{+}}$= -0.6 in the case of  
SM (Red-Solid), LHM (Green-Dotted), $E_6(\chi)$ (Blue-Dotted), ALRSM (Yellow-DotDashed)
and LRSM (Magenta-Dashed). 
The parameters used are $\Delta M$=100 MeV and $\theta=0.003$ in the case of $E_6$ and LR models, and
$f=1$ TeV and $c=0.3$ in the case of LHM..}}
\label{fig_angdistR}
\end{minipage}
\end{figure}

We next consider the angular distribution of one of
the secondary leptons. 
The way of calculating the angular distribution
is done in our earlier work \cite{APP}, which we 
follow here.  The angular distributions for 
different polarization combinations in case of different
models is calculated for $\sqrt{s}$ =800 GeV.
As in the earlier cases, the case of unpolarized beams and
the case of left-handed electron beams are not significant in the case
of angular distribution as well. 
But the case of right-handed polarization of electron beam along
with the left handed polarization of the positron beam provides much
better discrimination.  This distribution is shown in the Figs.~\ref{fig_angdist},~\ref{fig_angdistR}
for both the ideal and realistic degrees of beam polarizations. 
Qualitatively, the picture as regards discrimination between
the models, remains more or less the same as in the 
case of single 
energy distribution, except that there is a small difference 
between the two cases of 
left-right symmetric models considered. In the best case, the ALRSM model shows 
a deviation of around 45\% for the parameter set used.  

Fig.~\ref{fig_angdistR} shows the forward-backward asymmetric behaviour arising from the $Zee$ coupling,
which is different for different models considered. With this observation, we may obtain
the fraction of leptons emitted in the backward direction, which may be defined as
\[
f_{back}=\frac{\int_{-1}^0\left(d\sigma/d\cos\theta_l\right)~d\cos\theta_l}
{\int_{-1}^1\left(d\sigma/d\cos\theta_l\right)~d\cos\theta_l}.\]
This fraction is a useful quantity to consider 
in the case of unpolarized and left-polarized electron beams as well.  
In Table~\ref{tab:f_back} we present these fractions for 
$\sqrt{s}=500$ GeV, $800$ GeV and 1000 GeV. The result shows with 
unpolarized beams about 3-4 \% of the leptons are emitted in the
backward hemisphere.  The deviation is about 3\% in going from SM
to $E_6(\chi)$ and ALRSM which are more sensitive compared to other models.
At $\sqrt{s}$=800 GeV, the deviation becomes more significant and is about 8\%.
However LHM being more sensitive shows $\sim 36$ \% deviation at $\sqrt{s}$=500 GeV
which further increases with energy.  This above results are slightly
increased by switching on the beam polarization to left-handed electrons
and right-handed positrons.  The departure from ideal degrees of polarization does not play 
a significant role at the level of significant places retained in the
tables for this configuration.  
   
We see from Table~\ref{tab:f_back} that for the case of
polarization with about 90\% of right handed electrons and 60\% of left handed
positrons, the deviation increases. Owing to the near absence of the dominant 
t- channel neutrino diagram with this polarization configuration,
the new physics contribution shows up in the
fraction of the leptons emitted in the backward direction. 
Thus at $\sqrt{s}$= 800~GeV
for $E_6(\psi)$ and $E_6(\chi)$ there is about 15\% deviation from SM, whereas $E_6(\eta)$
is more constrained with only 5\% deviation. The symmetric models ALRSM and LRSM
are more sensitive with about 30\% deviation. 

In the case of completely right handed electrons
and left handed positrons, from the same table it may be
seen that all the models give rise to the same prediction
as the SM.  This is a consequence of the
near symmetric behaviour in the angular distributions as shown
in Fig.~\ref{fig_angdist}.
Due to this property the new physics effects in
$f_{back}$ and $A_{FB}$ are essentially wiped out.
It is important to note that this feature persists with
completely right-polarized electron and partially 
left-polarized positrons, and {\it vice versa}.

Another useful observable related to the angular asymmetry is the 
forward-backward asymmetry defined as
\begin{equation}
A_{FB}=\frac{\int_{-1}^0(d\sigma/d\cos\theta_l)~d\cos\theta_l-
             \int_{ 0}^1(d\sigma/d\cos\theta_l)~d\cos\theta_l}
            {\int_{-1}^1(d\sigma/d\cos\theta_l)~d\cos\theta_l}.
\end{equation}
The values for this asymmetry is also presented in Table \ref{tab:f_back}
for different models at three different collider energies.  This asymmetry
shows a similar behaviour as that of the fraction 
of the leptons emitted in the backward direction as regards
the beam polarization configurations and discrimination between
the models.  In fact, the two observables are intimately related:
 $A_{FB}=2f_{back}-1$.  However, since they are experimentally
realized differently, it may be appropriate to consider both
of them. 
\begin{table}
\begin{center}
\begin{tabular}{|c|c|c|c|c|c|c|c|c|} \hline
&&  &\multicolumn{2}{|c|}{}&\multicolumn{2}{|c|}{} &\multicolumn{2}{|c|}{}\\[2mm] 
& & &\multicolumn{2}{|c|}{$\sqrt{s}$ =500GeV}&
\multicolumn{2}{|c|}{$\sqrt{s}$ =800GeV}&
\multicolumn{2}{|c|}{$\sqrt{s}$ =1000GeV}\\[2mm] \cline{4-9}
&&&&&&&&\\
$P_{e^-}$ &$P_{e^+}$&Model
&$f_{back}$&$A_{FB}$ 
&$f_{back}$&$A_{FB}$
&$f_{back}$&$A_{FB}$\\[2mm] \hline\hline
  &  &SM      &0.0346&-0.9308 &0.0238 &-0.9524 &0.0208 &-0.9585 \\[1mm]    
  &  &$E_6(\chi)$      &0.0356&-0.9288&0.0257&-0.9487 &0.0234 &-0.9532           \\[1mm]  
 & &$E_6(\psi)$   &0.0347&-0.9305&0.0242&-0.9517 &0.0213& -0.9574              \\[1mm] 
0  &0  &$E_6(\eta)$       &0.0353 &-0.9293 &0.0252&-0.9496 &0.0228&-0.9545               \\ [1mm]
   &  &LRSM       &0.0347 &-0.9306 &0.0241&-0.9517 &0.0213& -0.9573              \\[1mm]
    &  &ALRSM       &0.0330 &-0.9339&0.0208&-0.9585 &0.0161&-0.9677               \\[1mm] 
      &  &LHM       &0.0474&-0.9053&0.0474&-0.9052 &0.0212&-0.9574               \\ [ 1mm] \hline
 &  &SM      &0.0324&-0.9353&0.0225&-0.9550 &0.0197&-0.9606 \\[1mm]     
  &  &$E_6(\chi)$      &0.0332&-0.9335&0.0241&-0.9518 &0.0219&-0.9561           \\[1mm]  
 & &$E_6(\psi)$   &0.0327&-0.9345&0.0232&-0.9537 &0.0206& -0.9588              \\[1mm]  
-0.9  & 0.6  &$E_6(\eta)$       &0.0332&-0.9336&0.0240&-0.9520 &0.0218&-0.9563               \\ [1mm] 
   &  &LRSM       &0.0328&-0.9343&0.0233&-0.9533 &0.0209&-0.9583               \\[1mm]  
    &  &ALRSM       &0.0312&-0.9376&0.0201&-0.9598 &0.0160&-0.9681               \\[1mm] 
      &  &LHM       &0.0440&-0.9120&0.0441&-0.9118 &0.0167&-0.9666               \\  [1 mm]\hline  
    &  &SM        & 0.1618  & -0.6765 & 0.1061 & -0.7878  & 0.0909 & -0.8183 \\[1mm]     
  &  &$E_6(\chi)$      &0.1694 & -0.6612 & 0.1208 & -0.7583  & 0.1116 & -0.7788           \\[1mm]  
 & &$E_6(\psi)$   & 0.1518 & -0.6964 & 0.0889 & -0.8222 & 0.0680 & -0.8639               \\[1mm]  
0.9  &-0.6  &$E_6(\eta)$       & 0.1587 & -0.6826 & 0.1013 & -0.7975  & 0.0845 & -0.8310               \\ [1mm] 
   &  &LRSM       & 0.1457 & -0.7087 & 0.0786 & -0.8427 & 0.0552 & -0.8896              \\[1mm]  
    &  &ALRSM       & 0.1426 & -0.7148 & 0.0648 & -0.8703  & 0.0268 & -0.9463               \\[1mm] 
      &  &LHM       & 0.2174 & -0.5653 & 0.2120 & -0.5760  & 0.2251 & -0.5498              \\  [1 mm]\hline  
  1   &-1  &All models        & 0.5834  & 0.1668  & 0.5392 & 0.0785   & 0.5264  & 0.0528  \\[1mm]  \hline

\end{tabular}
\end{center}
\caption{Fraction of leptons emitted 
in the backward direction, and the forward-backward asymmetry 
for all models for unpolarized and 
polarized beams with a parameter choice of $\theta$=0.003 and
$\Delta M$= 0.1 GeV. The parameters for LHM are $f$=1 TeV
and $c$=0.3}
\label{tab:f_back}
\end{table}

\subsection{Left-Right Asymmetry}

The left-right asymmetry is another important observable to be considered
for studying the new models.
We define the left-right asymmetry in the differential cross section as:
\begin{equation}
A^{diff}_{LR}=\frac{d\sigma(e^+_Re^-_L) / d\cos\theta-d\sigma(e^+_Le^-_R) / 
d\cos\theta}
{d\sigma(e^+_Re^-_L) / d\cos\theta+d\sigma(e^+_Le^-_R) / d\cos\theta},
\end{equation}
where $\theta$ is the $W$ scattering angle.

The $Z'$ of the $U(1)'$ models considered here couples to both left- and right- handed fermions,
but with varying relative couplings.  Thus, one would expect appreciable change in 
the asymmetry between the left- and right-polarized cross sections. But compared to these 
the new gauge boson in the LHM has the peculiar property
of coupling only to left-handed fermions as mentioned earlier.  

\begin{figure}[htb]
\includegraphics[width=6 cm,height=6 cm]{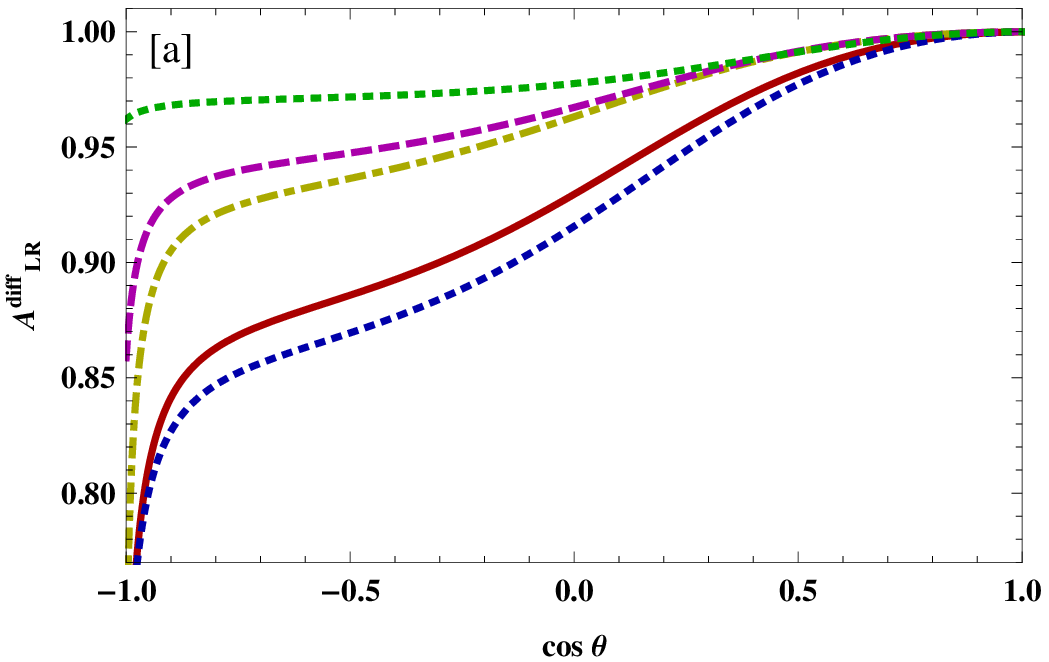}
\hspace{0.2cm}
\includegraphics[width=6 cm,height=6 cm]{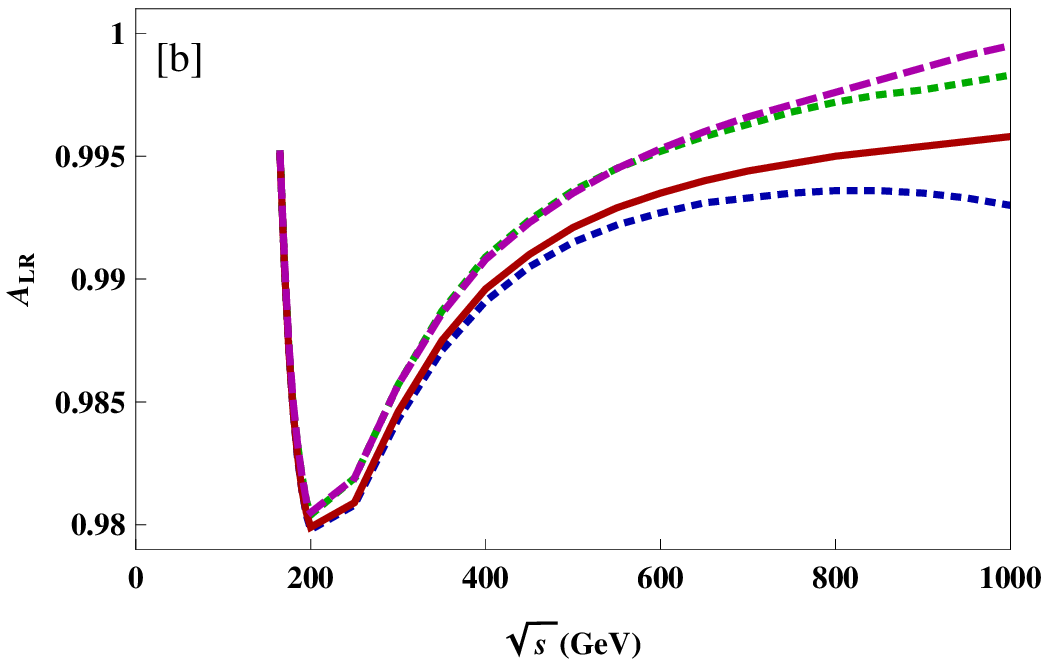}
\caption{$(a)$ Differential Left-right asymmetry as a function of the scattering angle for 
different models at $\sqrt{s}=800$ GeV, and $(b)$ the integrated left-right asymmetry as a function
of the centre of mass energy for different models: SM (Red-Solid), LHM(Green-Dotted), $E_6(\chi)$ (Blue-Dotted), LRSM(Magenta-Dashed)
and ALRSM (Yellow-DotDashed). Parameters used are $\theta=0.003$ and $\Delta M=100$ MeV for the $E_6$ and LR models, and $f=1$ TeV
and $c=0.3$ for the LHM.}
\label{fig:lr-asym}
\end{figure}

Fig.~\ref{fig:lr-asym}$(a)$ shows the LR asymmetry for $\sqrt{s}$=800 GeV.
Even at low energies the deviation becomes apparent. Notice also that the LHM now pair 
with the left-right symmetric models unlike in the case of energy and angular distributions 
shown in Figures ~\ref{fig_E} and \ref{fig_angdist}, providing us another tool for discriminating
different models from each other.

We may go one step further by considering an integral version of
this asymmetry as better efficiency may be obtained this way, 
by integrating each of the differential cross sections
from an opening angle $\theta_0$ up to an angle $\pi-\theta_0$,
for various realistic values of $\theta_0$ to which the data
can be integrated without difficulty.  
We define the integrated left-right asymmetry as:

\begin{equation}
A_{LR}=\frac{\sigma_{\theta_0}(e^+_Re^-_L\rightarrow WW)-
\sigma_{\theta_0}(e^+_Le^-_R\rightarrow WW)}
{\sigma_{\theta_0}(e^+_Re^-_L\rightarrow WW)+
\sigma_{\theta_0}(e^+_Le^-_R\rightarrow WW)}
\end{equation}
where $\sigma_{\theta_0}$ stands for $\int_{\theta_0}^{\pi-\theta_0}
\left(d\sigma/d\theta\right)~d\theta$.

This asymmetry, for different parameter models is plotted against the 
centre of mass energy in Fig.~\ref{fig:lr-asym}$(b)$ with $\theta_0$ =0. 
Dominance of the $t$-channel in the $W$-pair production establishes a 
highly forward peaked cross section, whereas as is seen from Fig.~\ref{fig:lr-asym}$(a)$ 
the deviations in LR asymmetry grows with the scattering angle, except for a region 
near $\theta =180$ degrees. Thus, the cut-off angle can be effectively used to increase
the deviation in the asymmetry. The results are presented in Table~\ref{tab:ALR}
for different cut off angles in different models at different centre of mass energies. An 
asymmetric cut off, meaning and integrated asymmetry between scattering angles, $\theta_{10}$
and $\pi-\theta_{20}$, with $\theta_{20}\neq \theta_{10}$, may improve the situation quantitatively.
Again, as in the case of other observables, our gaol in this report is to demonstrate the viability
of identifying deviations from the SM case, and further providing tools to disentangle different 
possible models. With this illustrative purpose in mind, we do not attempt to optimize the investigation.

\begin{tiny}
\begin{table}
\begin{center}
%\fontsize{26}{17}\selectfont Huge text
\begin{tabular}{|c|| c|| c c c|} \hline
    &  & &$A_{LR}$ & \\ \cline{3-5}
 Cut off angle  $\theta_0$ &Model &  $\sqrt{s}$=500GeV &  $\sqrt{s}$= 800 GeV &  $\sqrt{s}$= 1000 GeV\\[1mm] \hline
        &SM  &0.9921 &0.9950 &0.9958 \\ [0.6mm]
        &$E_6(\chi)$  &0.9914 &0.9936 &0.9930 \\ [0.6mm]
        &$E_6(\psi)$  &0.9930 &0.9967 &0.9984  \\ [0.6mm]
   0    &$E_6(\eta)$  &0.9924 &0.9956 &0.9968  \\ [0.6mm]
        & LRSM      &0.9935 &0.9976 &0.9995  \\ [0.6mm] 
        &ALRSM        &0.9935  &0.9976  &0.9995 \\ [0.6mm]
        &LHM       &0.9945  &0.9972  &0.9983  \\ \hline \hline
        &SM  &0.9853 &0.9872 &0.9877 \\ [0.6mm]
        &$E_6(\chi)$  &0.9842 &0.9836 &0.9800 \\ [0.6mm]
        &$E_6(\psi)$  &0.9870 &0.9916 &0.9955  \\ [0.6mm]
 15     &$E_6(\eta)$  &0.9860 &0.9889 &0.9909  \\ [0.6mm]
        & LRSM       &0.988 &0.9939 &0.9985  \\ [0.6mm]
        &ALRSM        &0.9879  &0.9938  &0.9985 \\ [0.6mm]
        &LHM       &0.9885  &0.9933  &0.9956  \\ \hline \hline       
        &SM  &0.9742 &0.9765 &0.9771 \\[0.6mm]
        &$E_6(\chi)$  &0.9723 &0.9703 &0.9638 \\ [0.6mm]
        &$E_6(\psi)$  &0.9772 &0.9846 &0.9917  \\ [0.6mm]
  30     &$E_6(\eta)$  &0.9755 &0.9797 &0.9834  \\ [0.6mm]
        & LRSM       &0.979 &0.9889 &0.9973  \\ [0.6mm]
        &ALRSM        &0.9787  &0.9885  &0.9971 \\ [0.6mm]
        &LHM       & 0.9808 &0.9890  &0.9928  \\ \hline  \hline 
          &SM  & 0.9601 &0.9622 &0.9628 \\ [0.6mm]
        &$E_6(\chi)$  &0.9574 &0.9531 &0.9434 \\ [0.6mm]
        &$E_6(\psi)$  &0.9648 &0.9753 &0.9866  \\ [0.6mm]
  45     &$E_6(\eta)$  & 0.9622 &0.9679 & 0.9738 \\ [0.6mm]
        & LRSM       & 0.9676 &0.9822  &0.9956  \\ [0.6mm]
        &ALRSM        & 0.9669 & 0.9811  & 0.9951 \\ [0.6mm]
        &LHM       & 0.9721 & 0.9845 & 0.9899 \\ \hline \hline
       &SM  & 0.9453 &0.9466 &0.9470 \\ [0.6mm]
       &$E_6(\chi)$  & 0.9418 &0.9348 &0.9224  \\ [0.6mm]
       &$E_6(\psi)$  & 0.9517 & 0.9652 & 0.9811 \\ [0.6mm]
  60   &$E_6(\eta)$    & 0.9483 & 0.9552 & 0.9638 \\ [0.6mm]
        &LRSM        & 0.9555 & 0.9750 & 0.9939 \\ [0.6mm]
        &ALRSM        & 0.9542 & 0.9727 & 0.9927 \\ [0.6mm]
         &LHM       & 0.9640 & 0.9807 & 0.9875 \\ \hline \hline
\end{tabular}
\end{center}
\caption{ Left-right asymmetry for different cut-off angles at
selected $\sqrt{s}$ values for different models considered
with a parameter choice of $\theta$=0.003 and
$\Delta M$= 0.1 GeV. The parameters for LHM are $f$=1 TeV
and $c$=0.3}
\label{tab:ALR}
\end{table} 
\end{tiny}

\section{Discussion and Conclusions} \label{Conclusions}
In the present work we have considered a class of additional $Z'$ models 
which are of interest to the linear collider community, both at the ILC as 
well as at CLIC.  While the masses of these bosons are already required to 
be significantly high, their imprint through mixing with the standard 
model $Z$ boson is the subject of this investigation.  By considering 
popular $E_6$ and left-right symmetric models like LRSM and ALRSM 
we have demonstrated that
the new physics signatures due to these models can be 
imprinted only at higher center of mass energies.
In the LHM model considered in our earlier work \cite{APP}, $Z'$ 
had the property of coupling only to left-handed fermions. 
This is in contrast to the other models considered here
where $Z'$ behaves like the SM $Z$. Thus compared to LHM, 
these models do not show appreciable deviation
at lower center of mass energies.   

While our focus in $W$-pair
production at ILC is on the unambiguous signal that it provides for $Z_{SM}-Z'$ mixing, we 
notice the interesting possibility of model discrimination here. 
For example, let us
compare the deviations of some of the observables from their SM values. In 
the case of energy and angular momentum distributions notice the qualitatively different behaviour
of left-right symmetric models compared to $E_6$ models and LHM.  Similarly, 
in the case of integrated and differential left-right asymmetries the $E_6$ model has a 
different qualitative behaviour compared to the LHM and left-right symmetric models.

Note that $WWZ_i~~(i=1,2)$ coupling is insensitive to differences between $Z'$ models, and therefore it is the
coupling of $Z'$ with the initial electron and positron that enables any possible model discrimination
in the present case. Thus, one would imagine that fermion pair production process is better suited 
to distinguish different models. In a recent study, the  
potential of the ILC to discriminate between $Z'$ models through fermion pair production is studied    
in Ref.~\cite{Osland1}. Here the sensitivity is studied by considering cross sections which 
are shown to be sensitive to new physics due to the availability of high 
statistics at the ILC.  It is also shown that beam polarization does 
enhance the sensitivity in an essential manner.
Similar studies with Drell-Yan dilepton production \cite{Osland2}
also probe distinguishability of different $Z'$ models at LHC. 
At the same time, our 
study indicates that $W$-pair production at ILC is capable
of supplementing the fermion pair production process in model discrimination
in the case of $Z'$ models.
More extensive analysis involving a parameter scan, also incorporating
a realistic collider-detector simulation is required to draw conclusions 
regarding ability of $W$-pair production in model discrimination at ILC, adapting the 
procedure in Ref.~\cite{Osland1}.  Alternatively one may contemplate adapting 
different asymmetries of the type we have proposed for fermion pair 
production as well. Perhaps a joint analysis involving both fermion pair production and $W$-pair production 
is needed to bring out the full potential of ILC in this regard.

Our conclusions for our models is as follows:
we have studied in detail each of the $E_6$ models denoted by
$\chi, \psi$ and $\eta$ and the LRSM and ALRSM models.
We find that at a centre of mass energy of 800 GeV, all these
models show a sharp deviation from the SM predictions for
energy-energy, single-energy and angular correlations of 
the decay lepton(s) generally for right-handed electron
and left-handed positron with realistic degrees of
polarization.   Curiously for ideal polarization the
decay lepton fraction and $A_{FB}$ are insensitive to
new physics for the same configuration above, as the
t- channel contribution is totally absent in this case.
The $\chi$, LRSM and ALRSM models 
are more sensitive than the other models even with unpolarized
beams.  The reason for the absence of sensitivity at lower energies
is due to the stringent bounds already present on the parameters
of the models.  The FB and LR asymmetries have also been
studied, with the latter being more sensitive to new physics
in all the models.  These models remain less sensitive than
the LHM model studied earlier.  Any indication of $U(1)'$ type of
$Z'$ in $W$-pair production at ILC is a clear signal of $Z_{SM}-Z'$
mixing. Apart from this, the present study points to the potential
of this process to supplement the fermion pair production
process in model discrimination, through a suitable combination of
observables considered.

%\texttt{Comparing our analysis with the
%work done in Ref.~\cite{Osland1} they have adopted $\chi^2$
%procedure to distinguish among the models for different values of $M_{Z'}$.
%They have found resolution regions where ILC can differentiate between the models
%for different parameter values.}  

Overall a strong polarization program at the ILC will
significantly enhance the diagnostic capability towards additional
$Z'$ bosons particularly at higher design energies.
%, and a suitable choice of different observable can discriminate different models (considered here) from each other.

\bigskip

\noindent
\textbf{Acknowledgments:} 
BA thanks the Department of Science and Technology,
Government of India and the Homi Bhabha Fellowships Council
for support during the course of these investigations. 
MP thanks the Department of Physics, Indian Institute of
Technology Guwahati for hospitality when part of this work was done.  PP
thanks BRNS, DAE, Government of India for 
support through a project ({\bf No.}: 2010/37P/49/BRNS/1446).

\end{document}